# LAPNet: Non-rigid Registration derived in k-space for Magnetic Resonance Imaging

Thomas Küstner, *Member, IEEE*, Jiazhen Pan, Haikun Qi, Gastao Cruz, Christopher Gilliam, *Member, IEEE*, Thierry Blu, *Fellow, IEEE*, Bin Yang, *Senior Member, IEEE*, Sergios Gatidis, René Botnar and Claudia Prieto, *Member, IEEE*

*Abstract*— Physiological motion, such as cardiac and respiratory motion, during Magnetic Resonance (MR) image acquisition can cause image artifacts. Motion correction techniques have been proposed to compensate for these types of motion during thoracic scans, relying on accurate motion estimation from undersampled motion-resolved reconstruction. A particular interest and challenge lie in the derivation of reliable non-rigid motion fields from the undersampled motion-resolved data. Motion estimation is usually formulated in image space via diffusion, parametric-spline, or optical flow methods. However, image-based registration can be impaired by remaining aliasing artifacts due to the undersampled motion-resolved reconstruction. In this work, we describe a formalism to perform non-rigid registration directly in the sampled Fourier space, i.e. k-space. We propose a deep-learning based approach to perform fast and accurate non-rigid registration from the undersampled k-space data. The basic working principle originates from the Local All-Pass (LAP) technique, a recently introduced optical flow-based registration. The proposed LAPNet is compared against traditional and deep learning image-based registrations and tested on fully-sampled and highly-accelerated (with two undersampling strategies) 3D respiratory motion-resolved MR images in a cohort of 40 patients with suspected liver or lung metastases and 25 healthy subjects. The proposed LAPNet provided consistent and superior performance to image-based approaches throughout different sampling trajectories and acceleration factors.

## I. INTRODUCTION

Magnetic resonance imaging (MRI) is a valuable and versatile tool in clinical diagnostics. Its capability of assessing anatomy and functional processes within the human body in a non-invasive manner makes it an essential imaging modality. However, MRI is prone to several artifacts which can deteriorate images significantly up to the point of non-diagnostic quality. Due to the long acquisition times in MRI, motion is one of the major extrinsic factors influencing image quality. Motion patterns can be categorized into rigid motion such as global translations or rotations of stiff structures which arises from movements of whole body parts and non-rigid motion including (local) deformations under an affine model which mainly occurs in the thorax and abdominal region caused by physiological motion. Other body parts, such as the bowel, can be affected as well. Patient and physiological motion induces ghosting along the phase-encoding direction and/or blurring of the image content. The manifestation in the image mainly depends on the imaging sequence and sampling trajectory.

Motion visualization, estimation and correction are thus important tasks when reconstructing or processing MRI data. Several prospective and retrospective motion compensation approaches have been developed to minimize or correct for motion induced artifacts. These include fast imaging sequences [1, 2] to enable acquisitions within resting periods (e.g. under breath-holds) or minimal motion (e.g. mid-diastole); tracking of motion by sensors (MR navigators [3-10], cameras [11], respiratory belts or electrocardiogram [12]) to limit data acquisition to periods with minimal movement (e.g. diastole, end-expiration); application of motion-robust acquisition schemes [13]; prospectively corrected acquisitions [14]; retrospective motion-corrected reconstruction [15, 16] and motion-resolved imaging [17, 18].

In relation to motion-corrected approaches, one can differentiate between the correction of rigid motion (e.g. head motion) and non-rigid periodic motion (e.g. respiratory and cardiac motion). Rigid motion can be tracked by MRI or other external sensors and modelled in k-space for translational motion as linear phase drifts that can be incorporated into the acquisition (prospective correction) or reconstruction (retrospective correction) schemes [8-10, 13]. Non-rigid motion can be tracked by MRI or sensors as well, but on the other hand is more challenging to correct as it involves local deformations in image space which are related to changes in the entirety of k-space (acquisition space) in a non-trivial way. Correction of non-rigid motion is therefore usually performed retrospectively on motion-resolved images [19]. Accordingly, a prior image reconstruction step is required.

T.K. and S.G. are with Medical Image And Data Analysis (MIDAS.lab), Department of Diagnostic and Interventional Radiology, Tübingen, Germany (e-mail: {thomas.kuestner | sergios.gatidis}@med.uni-tuebingen.de).
T.K, G.C. R.B. and C.P are with the School of Biomedical Engineering and Imaging Sciences, King's College London, St. Thomas' Hospital, London, United Kingdom (e-mail: {thomas.kuestner | gastao.cruz | rene.botnar | claudia.prieto}@kcl.ac.uk).
J.P and B.Y. are with the Institute of Signal Processing and System Theory, University of Stuttgart, Stuttgart, Germany (e-mail: st159861@stud.uni-stuttgart.de, bin.yang@iss.uni-stuttgart.de).
H.Q. is with the School of Biomedical Engineering, ShanghaiTech University, Shanghai, China (e-mail: qihk@shanghaitech.edu.cn).
C.G. is with Electronic and Telecommunications Eng, RMIT University, Melbourne, Australia (e-mail: christopher.gilliam@rmit.edu.au).
T.B. is with the Department of Electronic Engineering, Chinese University Hong Kong, Hong Kong, Hong Kong (e-mail: tblu@ee.cuhk.edu.hk).

To achieve fast acquisitions, motion-resolved thoracic and abdominal images are usually highly undersampled in k-space. In order to reconstruct aliasing-free images these methods rely on reconstruction schemes that for example exploit sparsity or low-rank redundancies in the spatial and/or motion directions to solve the ill-posed reconstruction problem [2, 20]. These approaches require careful parametrization and fine-tuning between regularization and data consistency to avoid residual aliasing (under-regularized) and staircasing or blurring artifacts (over-regularized).

After reconstruction, motion fields can be estimated in image space from the motion-resolved reconstructed images. Spatio-temporal redundancies can be exploited in the reconstruction to improve image quality [21-27] which implicitly perform a motion correction without directly relying on motion fields. Alternatively, motion-compensated reconstructions can be carried out which iterate between an image reconstruction and image registration step [28-35]. While motion-compensated methods have shown to efficiently utilize motion information, they require a significantly increased computational demand and the achievable imaging acceleration is limited by the quality of the motion-resolved undersampled reconstruction.

Non-rigid motion estimation is usually formulated in image space using diffusion-based [36], parametric spline-based [37] or optical flow-based registration methods [38]. In general, motion estimation can be guided by external motion surrogate signals [28, 39], initial motion field estimates [29, 30], from motion-aliased images [31] or low-frequency image contents, derived from central k-space data [40, 41]. It was shown in MR-MOTUS [41] that conventional image registrations can be modelled as a low-rank optimization for undersampled cases of non-rigid respiratory motion and rigid head motion..

Recently, deep-learning based approaches have been presented [42] to learn generalizable rigid [43, 44] and non-rigid [45-50] registrations in image space for medical images. FlowNet [51] and FlowNet-2 [52] have been proposed as supervised optical flow registration networks for fully-sampled 2D natural scene images and which has also been translated to registration of MR images [53, 54]. In CarMEN [55] motion fields were derived in an encoder-decoder convolutional neural network from dynamic multi-slice 2D MR image stacks to perform cardiac motion estimation.

To summarize the above, motion correction methods depend on i) reliable motion-resolved images or low-frequency k-spaces (i.e. low-resolution images) from which motion fields can be estimated, ii) require good initial motion field estimates, iii) rely on other external motion sensor signals, iv) constraint the imaging or v) motion field (e.g. local affine) optimization. In case of highly undersampled data, aliasing or blurring artifacts in the reconstructed images can impair the registration process as reconstruction errors can propagate into the image registration and/or low-resolution images may not provide sufficient information for accurate registration. Aliasing-free registration from accelerated acquisitions can be of use for i) integration into motion-compensated reconstructions, ii) inter- or intra-modality motion correction of other imaging acquisitions (e.g. PET/MR motion correction), or iii) investigation of subject-depending motion behaviour (e.g. cardiac wall motion).

In this work, we propose a deep-learning based approach to perform fast and accurate non-rigid motion estimation directly from acquired k-space based on optical flow equations. This work focuses on "image" registration carried out in k-space. The working principle originates from our recently introduced Local All-Pass (LAP) technique [56-58], an image-based 3D non-rigid optical flow registration. We will first describe the basic concepts of LAP and illustrate its extension to a k-space based 3D non-rigid registration [59]. We will then introduce the proposed deep-learning non-rigid registration network, named LAPNet, operating on motion-resolved k-space data to derive 3D deformation fields. The proposed LAPNet is compared against traditional image-based registrations (LAP and NiftyReg [60]) and FlowNet-S [51] operating on image inputs. We investigate the proposed approach in 40 patients with suspected liver or lung metastases and 25 healthy subjects for retrospectively and prospectively undersampled data of 3D respiratory motion-resolved MR imaging.

## II. THEORY

In 3D non-rigid image registration, one tries to estimate a motion field/deformation field/motion model $\underline{u} = [u_x, u_y, u_z]^T$ which maps all voxels of a moving image $\rho_m$ to a reference/fixed image $\rho_f$ by minimizing some dissimilarity metric $\mathcal{D}$

$$\hat{\underline{u}} = \arg\min_{\underline{u}} \mathcal{D}(\rho_f, \rho_m(\underline{u})). \qquad (1)$$

To solve this optimization task, one can utilize parametric (e.g. B-splines, thin-plate splines) [61] or non-parametric (e.g. optical flow, diffusion) [62] approaches. In the following, we will derive the notation for an efficient optical flow-based non-rigid registration in image space (II.A), which builds upon [56-58], from which we extend it to a non-rigid registration in k-space (II.B). A computationally efficient implementation is then achieved by deep learning registration network, named LAPNet (II.C).

### A. Local All-Pass (LAP) in image space

Under the assumption of local brightness consistency, i.e. the intensity remains constant when flowing from the moving to the fixed image, the volumetric images can be linked by a translation. The optical flow can be stated as

$$\rho_f(\underline{x}) = \rho_m(\underline{x} - \underline{u}(\underline{x})) \qquad (2)$$

for deforming a moving image $\rho_m \in \mathbb{R}^{N_x N_y N_z}$ to a fixed image $\rho_f \in \mathbb{R}^{N_x N_y N_z}$ with a translation $\underline{u}_t(\underline{x}) \in \mathbb{R}^3$ at voxel position $\underline{x} = [x, y, z]^T$ inside the field-of-view (FOV) determined by the 3D spatial size $N_x, N_y$ and $N_z$. This relationship also holds true for deforming local neighborhoods $\mathcal{W} \ll$ FOV with $\underline{x} \in \mathcal{W}$. Under the hypothesis of a motion flow continuum, one can approximate any global non-rigid flow $\underline{u} \in \mathbb{R}^3$ existing on a large grid by a large sum of many local translational flows $\underline{u}_t(\underline{x})$ in a smaller support $\mathcal{W}$,

$$\underline{u}(\underline{x}) = \sum_{\mathcal{W}} \underline{u}_t(\mathcal{W}) \times \mathcal{W}(x) \qquad (3)$$

In other words, any non-rigid deformation can be regarded as a sum of local translational displacements if 3D motion flow is smoothly varying.

For a translational flow, one can equivalently state the occurring translation in Fourier domain, following the shifting property of the Fourier transform

$$\rho_f(\underline{x}) = \rho_m(\underline{x} - \underline{u}_t) \iff v_f(\underline{k}) = v_m(\underline{k})e^{-j\underline{u}_t^T\underline{k}} \quad (4)$$

resulting in a multiplication of the moving k-space $v_m(\underline{k}) \in \mathbb{C}^{N_xN_yN_z}$ with a linear phase to obtain the fixed k-space $v_f(\underline{k}) \in \mathbb{C}^{N_xN_yN_z}$ at all k-space locations $\underline{k} = [k_x, k_y, k_z]^T$. Accordingly, if one then defines

$$H(\underline{k}) = e^{-j\underline{u}_t^T\underline{k}}, \quad (5)$$

the fixed k-space $v_f(\underline{k})$ becomes an all-pass (i.e. $|H(\underline{k})| = 1$) filtered version with $H(\underline{k})$ in Fourier domain of the moving k-space $v_m(\underline{k})$. Assuming ideal sampling with a sinc kernel, we have a digital version $H(\underline{k})$ of the all-pass filter with $(2\pi, 2\pi, 2\pi)$ periodic frequency response.

Putting together equations (3) to (5), one sees that solving for the non-rigid motion field $\hat{\underline{u}}$ in (1), i.e. performing a non-rigid registration (in image space), can be equivalently explained by finding the appropriate local all-pass filters in Fourier space and deriving the flow $\hat{\underline{u}}$ from it.

To estimate the all-pass filter $H(\underline{k})$, we use the key algorithmic idea proposed in [63], that the all-pass filtering between two images can be expressed linearly in terms of forward $F(\underline{k})$ and backward filtering $F(-\underline{k})$, yielding

$$H(\underline{k}) = e^{-j\underline{u}_t^T\underline{k}} = F(\underline{k})/F(-\underline{k}) \quad (6)$$

which is all-pass by design, i.e. $|H(\underline{k})| = |F(\underline{k})|/|F(-\underline{k})| = 1$. Consequently equation (4) respectively (2) becomes

$$\rho_f(\underline{x}) = h(\underline{x}) * \rho_m(\underline{x}) \iff f(-\underline{x}) * \rho_f(\underline{x}) = f(\underline{x}) * \rho_m(\underline{x}) \quad (7)$$

with all-pass $h(\underline{x})$, forward $f(\underline{x})$ and backward $f(-\underline{x})$ filters in image space. The discrete filters $f$ are approximated by a filter basis series $f_n(\underline{x})$

$$f(\underline{x}) = f_0(\underline{x}) + \sum_{n=1}^{N} c_n f_n(\underline{x}) \quad (8)$$

where $N$ denotes the number of optimal filter coefficients $c_n$ without loss of generality setting $c_0=1$. As pointed out in [64], a canonical basis $f_n$ with support in $\mathcal{W}$ of cubic size $W \times W \times W$ would act as an upper bound for the possible deformation of $W/2 - 1$. Hence, following the analysis in [65], an all-pass filter can be approximated by a finite filter basis $f_n$, if the basis spans the derivatives of an isotropic function.

The size of the filter basis $W$ confines the non-rigid motion estimation to estimating translational flows in a local region, respectively estimating local all-pass filters, given by the filter support $\mathcal{W}$ and defining then the LAP equation

$$f(-\underline{x}) * \rho_f(\underline{x}) - f(\underline{x}) * \rho_m(\underline{x}) = 0 \quad \forall \underline{x} \in \mathcal{W} \quad (9)$$

Combining (8) and (9) allows to conclude the non-rigid LAP registration problem in image space

$$\min_{\{c_n\}} \sum_{\underline{x} \in \mathcal{W}} \mathcal{D}\left[\mathcal{W}(\underline{x}) \cdot \left(f(\underline{x}) * \rho_m(\underline{x})\right), \right.$$
$$\left. \mathcal{W}(\underline{x}) \cdot \left(f(-\underline{x}) * \rho_f(\underline{x})\right)\right] \quad (10)$$
$$\text{s.t. } f(\underline{x}) = f_0(\underline{x}) + \sum_{n=1}^{N} c_n f_n(\underline{x}) \quad \forall \underline{x} \in \mathbb{R}^3$$

for which at each image position $\underline{x} \in \mathbb{R}^3$ the $N$ optimal filter coefficients $c_n$ are estimated by minimizing the dissimilarity $\mathcal{D}$ (e.g. mean-squared-error, MSE) between $\rho_m$ and $\rho_f$. For sake of simplicity, $\mathcal{W}$ describes the window function indicator of the neighborhood $\mathcal{W}$ in (3). The local all-pass filter yields the local translational flow at the central voxel in $\mathcal{W}$ and can be derived from (6) as

$$\underline{u}_t = j\frac{\partial \ln H(\underline{k})}{\partial \underline{k}}\bigg|_{\underline{k}=\underline{0}}$$
$$\iff \underline{u}_t = 2\left[\frac{\sum_{\underline{x}} xf(\underline{x})}{\sum_{\underline{x}} f(\underline{x})}, \frac{\sum_{\underline{x}} yf(\underline{x})}{\sum_{\underline{x}} f(\underline{x})}, \frac{\sum_{\underline{x}} zf(\underline{x})}{\sum_{\underline{x}} f(\underline{x})}\right]^T. \quad (11)$$

Sliding the window $\mathcal{W}(\underline{x})$ over all voxel positions provides the non-rigid flow $\underline{u}$ as stated in (3).

In order to deal with motion of varying strength, a multi-resolution approach is applied in which the size of $\mathcal{W}$ is decreased per multi-resolution step, i.e. coarse-to-fine estimation.

### B. Non-rigid registration in k-space

Following the key idea of LAP that any non-rigid deformation can be regarded as local translational displacements and together with the Fourier shift property, a k-space representation is obtained in (4) linking a moving k-space $v_m(\underline{k}) = \mathcal{F}\rho_m(\underline{x})$ to a fixed k-space $v_f(\underline{k}) = \mathcal{F}\rho_f(\underline{x})$ at all k-space locations $\underline{k}$, where $\mathcal{F}$ describes the Fourier transform.

In order to carry out the estimation of local translations in Fourier domain (i.e. k-space), we need to consider the local windowing $\mathcal{W}$

$$\rho_\mathcal{W}(\underline{x}) = \mathcal{W}(\underline{x}) \cdot \rho(\underline{x}) \iff v_\mathcal{W}(\underline{k}) = T(\underline{k}) * v(\underline{k}) \quad (12)$$

which corresponds in k-space to the convolution by a phase-modulated (for various $\underline{x}$ positions) tapering function $T(\underline{k})$. Consequently, transforming (10) in Fourier domain

$$\min_{\{c_n\}} \sum_{\underline{k} \in \mathbb{R}^3} \mathcal{D}\left(T(\underline{k}) * \left(F(\underline{k})v_m(\underline{k})\right), T(\underline{k})\right.$$
$$\left. * \left(F(-\underline{k}) v_f(\underline{k})\right)\right) \quad (13)$$
$$\text{s.t. } F(\underline{k}) = F_0(\underline{k}) + \sum_{n=1}^{N} c_n F_n(k) \quad \forall \underline{k} \in \mathbb{R}^3$$

yields the non-rigid k-space registration based on the LAP optical flow concept. However, not for all dissimilarity $\mathcal{D}$ in image space exists a representable counterpart in Fourier domain. Following the Parseval equation we restrict ourselves to $\mathcal{D}$ being the MSE.

Please also note that in the k-space version, summation is required over all $\underline{k}$ positions and for shifted tapering supports at all $\underline{x}$ positions. Hence, carrying out a registration over the complete spectral support of the FOV $\in \mathbb{R}^{N_x \times N_y \times N_z}$ requires $\mathcal{O}\left(\left(N_x N_y N_z\right)^{12}\right)$ operations at each iteration which can be computationally demanding. In comparison, the image-based version only operates on the smaller window $W \ll \text{FOV}$, yielding $\mathcal{O}(W^6 N_x N_y N_z)$ operations. It should be noted that for undersampled acquisitions, only the acquired k-space locations need to be visited in the registration. Hence for an undersampled acquisition with an acceleration factor of $R$, we obtain a computational reduction of $R^6$ which reduces computational burden only slightly.

We therefore seek to simplify these operations and significantly reduce the computational burden by learning an appropriate registration network that can carry out the previously described non-rigid registration in k-space and which is furthermore not prone to aliasing artifacts arising from undersampled acquisitions, i.e. generalizes for different acceleration strategies.

### C. LAPNet: Non-rigid registration network in k-space

First, we will have a closer look at the effect of the tapering function $T$. In image domain, the windowing $\mathcal{W}$ nulls the image outside of the chosen filter support and allows thus to crop the image to the smaller cubic size. A similar operation can be achieved in k-space by tapering with $T$ followed by a regridding to yield an input cube of size $W \times W \times W$ with $W \ll N_{x,y,z}$. This minimizes the required input dimensions of the network and thereby the memory footprint and number of trainable parameters in the network.

The proposed deep-learning architecture, denoted as LAPNet, for non-rigid registration in k-space is depicted in Fig. 1. The zero-filled 3D moving $v_m$ and fixed $v_f$ k-space is convolved with the tapering function $T$ following (12) and subsequently regridded to an empirically optimized size of 33x33x33.k-Spaces are sliced along one spatial dimension and real and imaginary parts of the moving and fixed k-space are concatenated along the channel direction. This bundle of tapered k-space patches of size $33^2$x4 is then passed through a succession of convolutional filters with dyadic increase in kernels and leaky ReLU activation function. Kernel sizes are depicted in Fig. 1. In the last layer a fully connected regression is performed on the average pooled feature map to estimate the in-plane deformations $\hat{u}_{1i}, \hat{u}_{2i}$ at the given central location of the k-space input patch for the $i$-th run. To obtain a 3D deformation field $\underline{\hat{u}} = [\hat{u}_x, \hat{u}_y, \hat{u}_z]$, the registration is performed on two orthogonal spatial directions. In the first run ($i = 1$), k-spaces are sliced along the last spatial dimension (readout direction), yielding $\hat{u}_{11}, \hat{u}_{21}$. In the second run, slicing can be done e.g. along the first dimension for processing this bundle of k-space patches yielding $\hat{u}_{12}, \hat{u}_{22}$. Results are then merged with the previous run to yield $\hat{u}_x = \hat{u}_{11}, \hat{u}_y = 0.5(\hat{u}_{21} + \hat{u}_{12}), \hat{u}_z = \hat{u}_{22}$ at the central voxel location of the k-space cube $\mathcal{W}$. The whole non-rigid deformation field $\underline{\hat{u}}$ (in image domain) is obtained by estimating the deformations $\hat{u}_x, \hat{u}_y, \hat{u}_z$ at all voxel locations by processing tapered k-spaces patches sequentially, i.e. sliding the window $\mathcal{W}$ over all voxels (as indicated by the arrows in Fig. 1). Flows are estimated from each k-space patch and correspond to the voxel location of the patch center. Flows are in image domain, due to the last fully-connected layer learning a weighted transformation.

This non-rigid registration follows the principle optimization stated in (13) where the network learns the canonical basis of (all-pass) filters (Fig. 1 convolution filters) as expressed in (8) for a local neighborhood (Fig. 1 tapering) as described in (12). The $N = 6$ consecutive convolutional filters can be regarded as learnable coefficients and basis functions $c_n \hat{f}_n$. Training is performed in a supervised manner to optical-flow derived diffeomorphic reference flows of (11). The estimated flows by LAPNet can thus be assumed to be diffeomorphic as well. Moreover, performing flow estimation on a k-space input patch follows the idea of approximating a global non-rigid flow by local translational deformations.

The network is trained in a supervised manner on pairs of moving $v_m$ and fixed k-space $v_f$ inputs with the corresponding reference motion field $\underline{u}_{\text{Ref}}$ derived from the image-based LAP [56, 57] (details described in section III.B). The squared end-point error (sEPE)

$$\text{sEPE} = L_{\text{LAPNet}} = \sum_{i \in \{x,y,z\}} \left(u_{\text{Ref},i} - \hat{u}_i\right)^2 \quad (14)$$

was employed as the training loss. The network resulted in ~25 million trainable parameters and was trained by an Adam optimizer [66] ($\beta_1 = 0.9, \beta_2 = 0.999$) for an initial learning rate of $2.5 \cdot 10^{-4}$ with learning rate scheduler (division by 2 every 50000 iterations) and a batch size of 64 over 50 epochs on a Nvidia Titan RTX GPU. Training and test data are further specified in section III.B.

The source code is publicly available under MIT license: github.com/lab-midas/lapnet

## III. METHODS

### A. In-vivo 4D MR acquisition

3D motion-resolved k-space data was obtained in a cohort of 40 patients (60 ± 9 years, 22 female) with suspected liver or lung metastases [58] and 25 healthy subjects (31 ± 4 years, 10 female) [18]. The study was approved by the local ethics committee and all subjects gave written consent. Imaging was performed on a 3T PET/MR (Biograph mMR, Siemens Healthcare, Erlangen, Germany) equipped with a phased array body and spine coil. A 3D T1 weighted spoiled gradient echo sequence was acquired in coronal orientation with a continuous variable-density Poisson Disc undersampling [67] for a time of acquisition (TA) of 5 min. The remaining imaging parameters were TE = 1.23ms, TR = 2.60ms, bandwidth = 890Hz/px and a flip angle of 7°. A matrix size of $N_x \times N_y \times N_z$ = 256 x 256 x 144 (RO x PE x 3D ⇔ FH x LR x AP) was acquired covering a field-of-view of 500 x 500 x 360 mm$^3$. A 2D MR self-navigation signal (256 x 8 x 1, RO x PE x 3D) was acquired each 200 ms serving as gating signal [18]. MR data were retrospectively gated into $N_t = 8$ respiratory bins, ranging from end-expiratory to end-inspiratory position, with a Gaussian view-sharing amongst neighbouring bins. Coil sensitivity maps were determined by ESPIRiT [68] from the time-averaged (across all respiratory gates) central fully sampled k-space data. Images were reconstructed by a FOCal Underdetermined System Solver (FOCUSS) [67] producing motion-resolved and complex-valued target images $\rho_{\text{target}} \in \mathbb{C}^{N_x N_y N_z N_t}$ respectively target k-spaces $v_{\text{target}} = \mathcal{F}\rho_{\text{target}} \in \mathbb{C}^{N_x N_y N_z N_t}$.

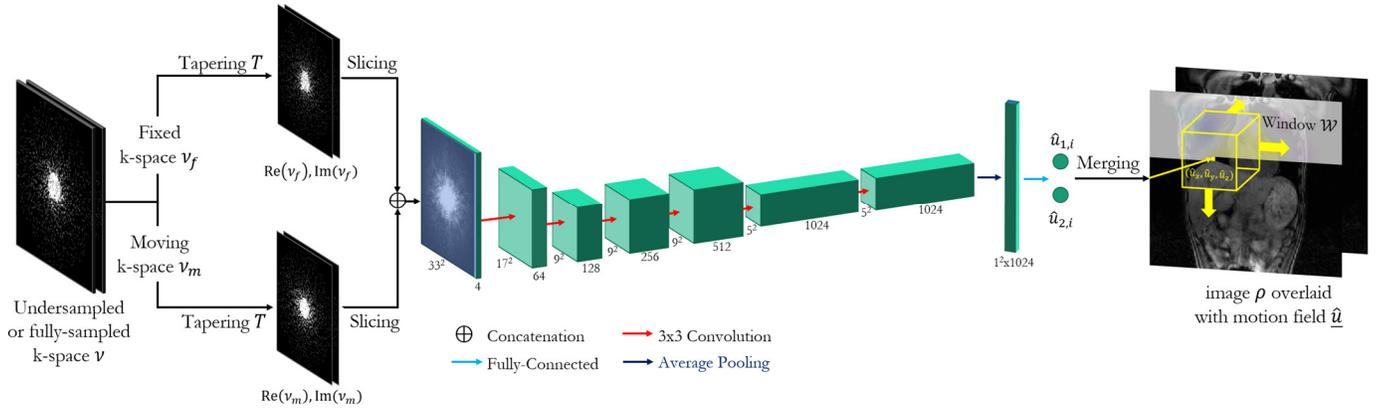

Fig. 1: Proposed LAPNet to perform non-rigid registration in k-space. Moving $v_m$ and fixed $v_f$ k-spaces are tapered to a smaller support. Slicing along one dimension and concatenating the respective real and imaginary parts yields the 2D input to the network. The bundle of k-space patches (with four input channels) is processed in a succession of convolutional filters (kernel sizes and channels are stated) with dyadic increase in channels to estimate the in-plane flows $\hat{u}_{1i}, \hat{u}_{2i}$ of the $i$-th run at the central voxel location determined by the tapering $T$ respective window $\mathcal{W}$ for size $W \times W \times W$ with $W = 33$. A 3D deformation field $\hat{\underline{u}} = [\hat{u}_x, \hat{u}_y, \hat{u}_z]$ is obtained by merging estimations from two runs on orthogonal spatial directions and for every voxel in a sliding window procession (yellow arrows).

### B. Training and test data

Training data was derived from reconstructed motion-resolved target data $\rho_{\text{target}}/v_{\text{target}}$. Reference flows $\underline{u}_{\text{Ref}}$ in training were composed of *real*, *smooth* and *augmented* flows. Motion fields $\underline{u}_{\text{real}}$ were obtained with the image-based LAP registration for $N = 4$ over 5 multi-resolution levels with window sizes per level of $W = 64, 32, 16, 8, 4$ between end-expiratory fixed $\rho_f = \rho_{\text{target}}(t = 1)$ and remaining moving bins $\rho_m = \rho_{\text{target}}(t > 1)$ of the target image (denoted as *real* as described realistic motion) with $t \in [1, N_t]$ being the bin number. Moreover, arbitrary smoothly varying flows/motion fields $\underline{u}_{\text{smooth}}$ were generated with a maximum displacement of 10 voxels drawn from multivariate Gaussian distribution [56, 57] (denoted as *smooth*). Flows $\underline{u}_{\text{real}}$ were randomly augmented by smoothing (convolution with 5x5x5 Gaussian low-pass), translating (-10 to 10 voxels in each direction), rotating (-25° to 25° in all planes) and by multiplication with arbitrary smoothly varying flows (denoted as *augmented*).

One objective of the proposed approach is to provide non-rigid registration from undersampled datasets without the need of image reconstruction, i.e. not being limited by aliasing artifacts. In training, the fixed $v_f$ and moving $v_m$ k-spaces are retrospectively undersampled with a sampling operator $\phi$. Two undersampling strategies $\phi \in \mathbb{R}^{N_x N_y N_z}$ are employed: 1) a variable-density Poisson-Disc undersampling (*vdPD*), reflecting a typical incoherent undersampling for compressed sensing like reconstructions and 2) taking a fully-sampled elliptical central region (*center*), reflecting a low-resolution acquisition. Acceleration factors were randomly distributed in the range of $R = 1$ to 30 where $R = 1$ corresponds to the *fully-sampled* case, i.e. $v_f = \phi v_{\text{target}}(t = 1)$ with $\phi(\underline{k}) = 1 \, \forall \underline{k}$. LAPNet was trained jointly on the different undersampling strategies and acceleration factors.

For training, the fixed input corresponds to the *fully-sampled* ($R = 1$) or *vdPD/center* ($R > 1$) undersampled k-space $v_f = \phi v_{\text{target}}(t = 1)$ of the target acquisition in end-expiratory position and the moving input relates to the registered *fully-sampled* or *vdPD/center* undersampled k-space $v_m = \phi \mathcal{F} \rho_{\text{target}}(\underline{x} \mp \underline{u}_{\text{Ref}}, t = 1)$ deformed with the forward motion model $\underline{u}_{\text{Ref}}$ (*real*, *smooth* or *augmented*) and bilinear interpolation.

Training data was created by taking k-space patches together with their corresponding flows at random spatial locations in coronal and sagittal orientation from 33 patients and 18 healthy subjects. A total of 15 million training samples with different types of motion (*real*, *smooth*, *augmented*) and undersampling (*fully-sampled*, *vdPD*, *center*) were generated approximately every 5th epoch with empirically determined ratios of 40% *real*, 20% *smooth* and 40% *augmented*. For training a stride of 1 was performed for the Window $\mathcal{W}$.

For testing, k-space patches of 7 patients and 7 healthy subjects (not included in training) were taken and flow estimation was conducted at every 2nd k-space location for LAPNet, i.e. a sliding window with a stride of 2 was performed to boost registration performance during inference, which showed best performance without loss of accuracy. In the *fully-sampled* case ($R = 1$) fixed and moving k-space are extracted from $v_{\text{target}}$. In testing for $R > 1$, the continuous sampling in the acquisition is cropped to a shorter scan time duration, and thus mimicking a prospective undersampling. Undersampled fixed and moving k-space inputs for testing were taken after respiratory motion binning of this prospectively accelerated acquisition.

### C. Experiments

For comparison, two conventional image-based 3D non-rigid registrations were performed by the image-based LAP (denoted as imageLAP) [56-58] which is an optical flow method and by NiftyReg [60] which used a free-form deformation cubic B-spline model [69]. Please note that for $R = 1$, the imageLAP provides the reference flow used in training the networks. Comparison to imageLAP is therefore only performed for $R > 1$. Conventional methods serve as comparison to investigate impact of image acceleration on image-based and non-data adaptive registrations. Both conventional methods were run on an Intel Xeon E5-2697 CPU. Experiments on the conventional non-rigid registration in k-space can be found in [59].

For further comparison, an image-based FlowNet-S [52] network was trained on the respective *fully-sampled* ($R = 1$) or *vdPD/center* ($R > 1$) undersampled zero-filled images and corresponding flows $\underline{u}_{\text{Ref}}$. Full 2D coronal and sagittal images are used as input. The network was trained by an Adam optimizer ($\beta_1 = 0.9, \beta_2 = 0.999$) for an initial learning rate of $10^{-4}$ with learning rate scheduler (division by 2 every 5000 iterations) and a batch size of 16 over 50 epochs to minimize the sEPE loss in (14). Remaining hyperparameters were optimized but found to coincide with published parameters [51]. Training database composition differed to LAPNet in the sense that the best combination of undersampling (*fully-sampled*, *vdPD*, *center*) was empirically optimized to yield the smallest possible validation loss. Two instances of FlowNet-S were hence trained for i) *fully-sampled + vdPD* and ii) *fully-sampled + center*. If not stated otherwise, the best performing case is reported.

The two different undersampling strategies enable to infer which k-space locations can potentially contribute to the non-rigid registration, i.e. is it beneficial to include high-frequency components (*vdPD*) or is it better to only focus on the low-frequency range (*center*). The latter undersampling case thereby also allows a comparison to previously published methods [40, 41] operating on low-resolution images (obtained from central k-space sampling) or central k-space data. Testing was performed considering the respective undersampling strategies (*fully-sampled*, *vdPD*, *center*) independently, for LAPNet and FlowNet-S.

### D. Evaluations

The end-point error $\text{EPE} = \left\| \underline{\hat{u}} - \underline{u}_{\text{Ref}} \right\|_2$ and end-angulation error $\text{EAE} = \arg(\underline{\hat{u}}, \underline{u}_{\text{Ref}})$ between the estimated motion field $\underline{\hat{u}}$ was compared with the reference motion field $\underline{u}_{\text{Ref}}$ obtained from imageLAP ($R = 1$) of the target acquisition over the whole FOV. Structural similarity index (SSIM) [70], normalized root mean squared error $\text{(NRMSE)} = 1/N \sqrt{\text{MSE}}$, peak signal-to-noise ratio $\text{(PSNR)} = 10 \log_{10} 1/|\text{MSE}|$ and normalized cross-correlation (NCC) were calculated between the deformed moving image $\rho_d = \rho_{\text{target}}(\underline{x} - \underline{\hat{u}}, t > 1)$ and the end-expiratory fixed target $\rho_f = \rho_{\text{target}}(t = 1)$. Bilinear interpolation was performed for non-integer grid points. All quantitative results are reported as mean ± one standard deviation over all voxel positions, gates and test subjects. Due to the lack of a gold-standard ground-truth motion, quantitative results can only be interpreted in relation to each other. Statistical significance was determined with a paired Welch's t-test and Bonferroni correction under the null hypothesis of equal means for unequal variances. P-values < 0.05 were considered statistically significant.

### IV. Results

The proposed LAPNet showed statistically significant superior performance in the quantitative analysis of EPE and EAE over changing acceleration factors for *vdPD* and *center* sampling as depicted in Fig. 2 and Supplementary Table I. LAPNet has the lowest EPE and EAE amongst all cases and shows a consistent performance throughout acceleration factors. Deviations amongst subjects in conventional image-based methods (imageLAP, NiftyReg) are larger indicating a less consistent performance. *Center* sampling shows slightly higher EPE and EAE metrics towards higher accelerations than *vdPD* indicating that high-frequency information is beneficial for the registration task (both in k-space and image space). The proposed k-space based LAPNet is less affected by this information loss and generalizes better.

Quantitative similarities between the registered images and the fixed images by a voxel-wise intensity comparison is shown as violin distributions in Fig. 6 over all test subjects and accelerations. Percentage differences and statistical pairwise testing to the proposed LAPNet are summarized in Supplementary Table I. The proposed LAPNet outperformed all other methods in each respective metric with statistical significance. LAPNet showed a more consistent performance with lower standard deviations. In *center* sampling slightly improved performance of image-based methods were observed.

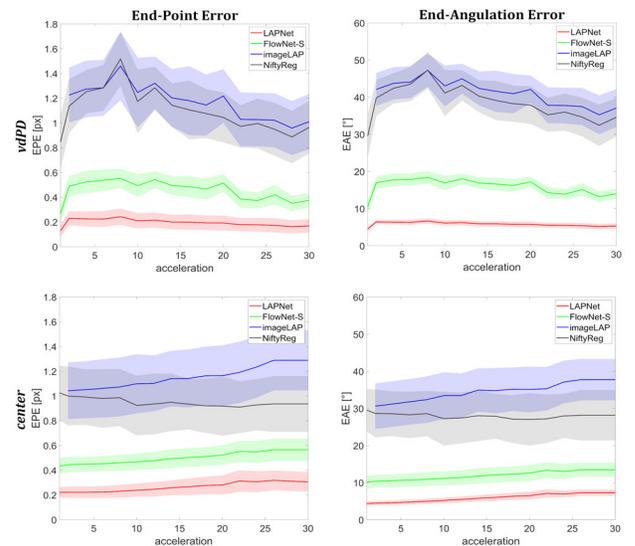

Fig. 2: End-point error (EPE) and end-angulation error (EAE) over all test subjects and gates for comparison of reference motion (*fully-sampled* imageLAP) to the proposed LAPNet in k-space, image-based FlowNet-S, image-based LAP (imageLAP) and image-based NiftyReg. Mean (solid line) and standard deviation (shaded area) are depicted for changing acceleration factors in *vdPD* and *center* sampling. Please note that imageLAP starts at acceleration $R = 2$ as it serves as reference for $R = 1$.

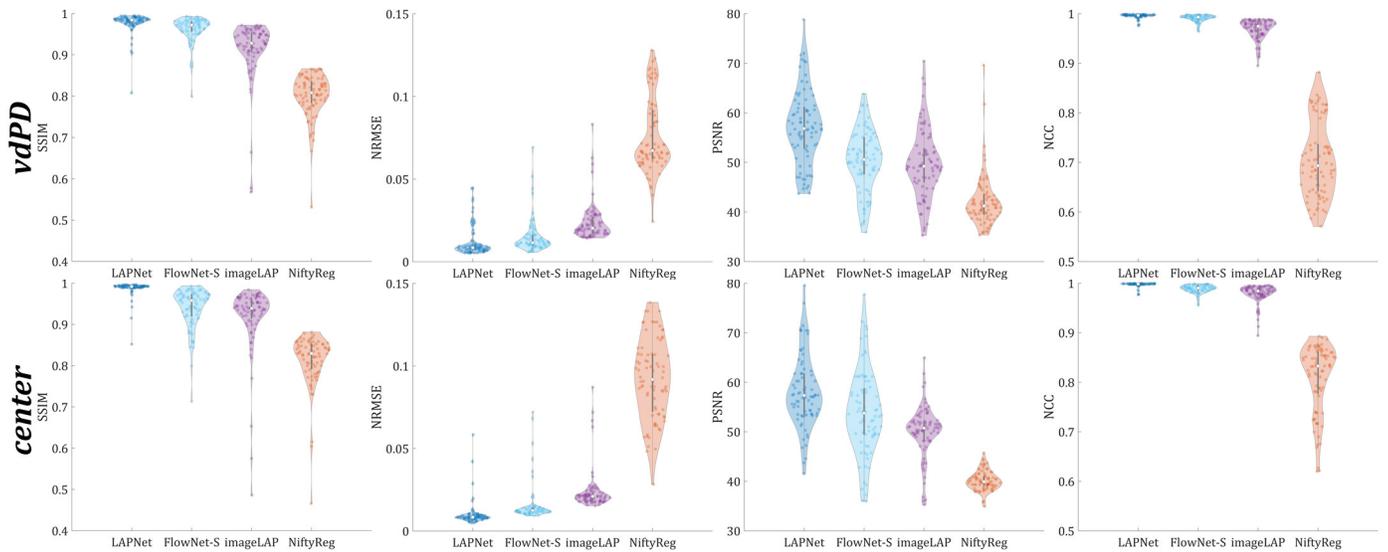

Fig. 3: Quantitative evaluation between registered images and end-expiratory target over all accelerations and gates by means of structural similarity index (SSIM), normalized root normalized squared (NRMSE), peak signal-to-noise ratio (PSNR) and normalized cross-correlation (NCC) for the proposed LAPNet in k-space, image-based FlowNet-S, image-based LAP (imageLAP) and image-based NiftyReg. Mean (central white dot) and standard deviation (vertical gray bar) are depicted in both undersampling strategies (*vdPD* and *center*).

The respiratory motion-resolved images of a representative patient with a neuroendocrine tumor in the liver are depicted in Fig. 4 for a *fully-sampled* acquisition and for prospectively undersampled *vdPD* and *center* acquisitions with 8x and 30x acceleration. Deformation fields were obtained by the proposed LAPNet, FlowNet-S, imageLAP and NiftyReg. The LAPNet (i.e. k-space non-rigid registration) shows good performance over all acceleration factors with close resemblance to the target motion field. Motion flows are only estimated in the body trunk with LAPNet whereas estimation in the static (noisy) background is relatively low or smoothed out in comparison to the conventional methods (imageLAP and NiftyReg).

Learning-based approaches (LAPNet and FlowNet-S) better generalize to the changing input data (remaining aliasing and noise impact) than the conventional methods. The data-adaptive learning allows to generalize a network-based registration to a certain extent for different acceleration factors, i.e. a network is able to see through the aliasing and noise artifacts. However, one can appreciate the performance difference between LAPNet (k-space registration) and FlowNet-S (image-space registration). The LAPNet is less obstructed by aliasing artifacts, resulting in smoother and more consistent flows amongst accelerations with closer agreement to the actual underlying respiratory motion which is also reflected in this subject in a statistically significant lower ($p < 0.001$) EPE of 64% ± 3% and lower EAE of 42% ± 6% in LAPNet than FlowNet-S. It can also be appreciated that for higher accelerations the aliasing is less pronounced in the images. Images are blurrier and smoothed out which arises from the fact that *vdPD* sampling is concentrated towards the low-frequency central region similar to the *center* sampling.

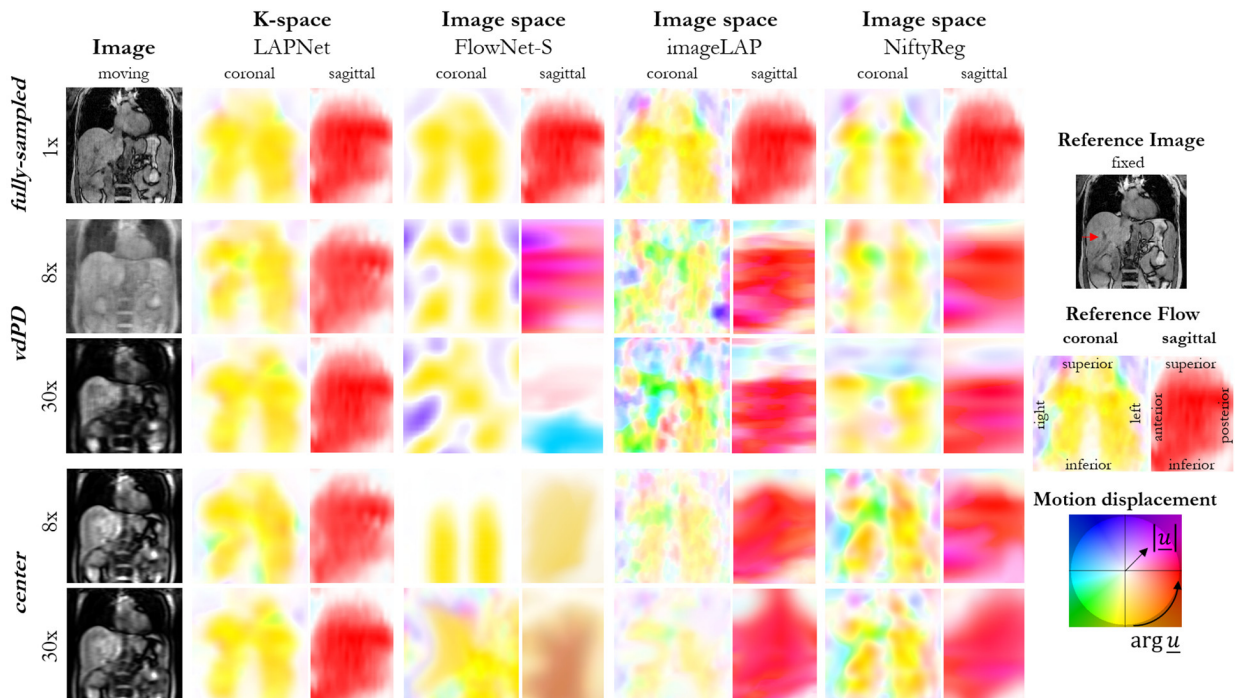

Fig. 4: Respiratory non-rigid motion estimation in a patient with neuroendocrine tumor in the liver (pointed out by red arrow) by the proposed LAPNet in k-space in comparison to image-based non-rigid registration by FlowNet-S (deep learning), image-based LAP (imageLAP; optical flow) and NiftyReg (cubic B-Splines). Estimated flow displacement are depicted in coronal and sagittal orientation. Reference flows are obtained from imageLAP on fully-sampled images. Undersampling was performed prospectively with a *vdPD* and *center* undersampling for 8x and 30x acceleration. Super-inferior liver dome displacement between end-expiratory and end-inspiratory of 1.5 cm was observed.

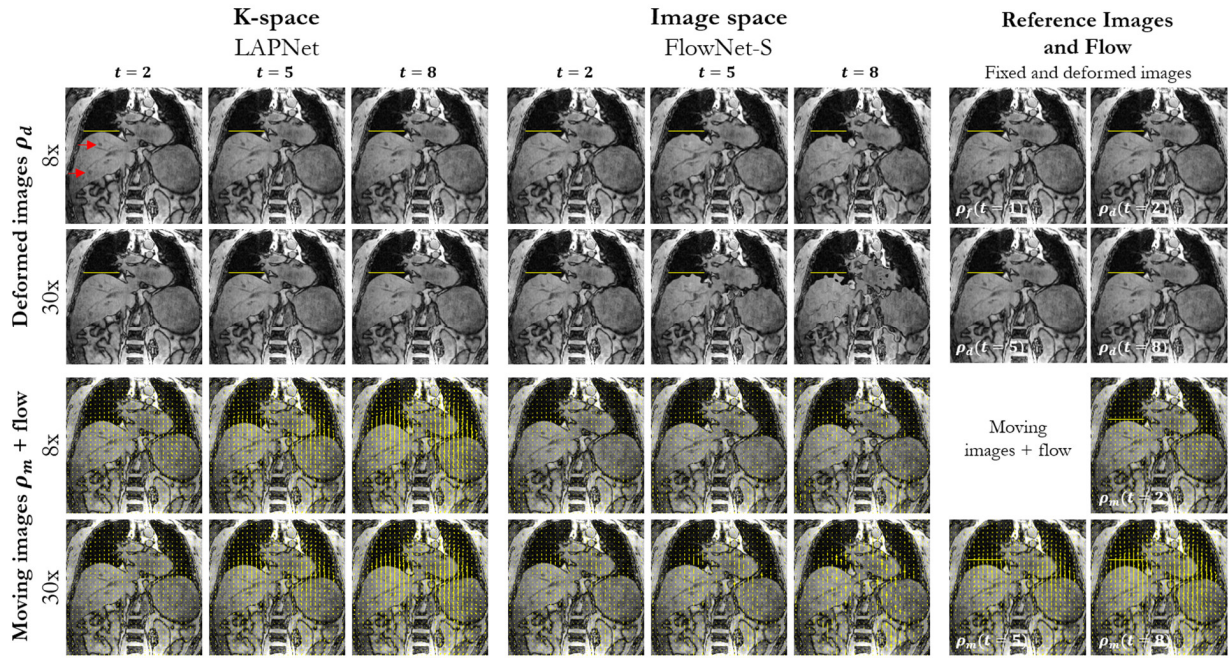

Fig. 5: Respiratory non-rigid motion estimation in a patient with pancreatic carcinoma and liver metastasis (pointed out by red arrows). Motion displacement is estimated by the proposed LAPNet in k-space in comparison to image-based non-rigid registration of FlowNet-S for *vdPD* undersampling with 8x and 30x acceleration. *Fully-sampled* moving images overlaid with estimated flows and deformed images are shown. Reference images and flows depict the image-based imageLAP registration. Horizontal yellow lines indicate liver dome displacement. Super-inferior liver dome displacement between end-expiratory (t=1) and end-inspiratory (t=8) of 2.8 cm was observed.

LAPNet produced consistent results over all acceleration factors as well as for different sampling strategies. Image-based methods (FlowNet-S, imageLAP and NiftyReg) show a similar trend for reducing performance with increasing acceleration factors. FlowNet-S performs better than conventional methods but shows substantial misregistration at the diaphragm. Misregistered flows in the image background were increased for FlowNet-S with *center* and *vdPD* sampling even for carefully optimized training database composition. FlowNet-S handles low-resolution image input (*center*) statistically significant better ($p < 0.001$) than undersampling artifact-affected images (*vdPD*) for accelerations $R \leq 15$. Registration of conventional image-based methods (imageLAP and NiftyReg) failed for *center* with $R > 10$, yielding random flows. Overall letting us conclude that high-frequency samples, i.e. image edges, carry relevant information for the image-based registrations which are inevitable lost at one point for a decreasing sampled central region size (increasing acceleration $R$), respectively stronger low-pass filtered image.

In the coronal orientation, superior-inferior respiratory motion is mainly dominating. Static non-moving regions such as the spine are not deformed by the proposed LAPNet, as well as the image-based registrations for the *fully-sampled* case, indicating that local non-rigid deformations can be correctly captured. In sagittal orientation, the superior-inferior respiratory motion shows an additional posterior displacement.

Fig. 5 shows the deformed images obtained from non-rigid registration in undersampled data for the proposed LAPNet and FlowNet-S in a patient with pancreatic carcinoma and liver metastasis. For increasing acceleration factors, a rising trend of misregistered local deformations and non-smooth flows were obtained with the image-based methods yielding overall a less consistent performance. The proposed LAPNet provided consistently good flow estimation over all motion states (end-expiratory to end-inspiratory) and accelerations. Deformed images of LAPNet matched the fixed reference image better than that of FlowNet-S. In general, performance of FlowNet-S was inferior to LAPNet.

Fig. 6 and 7 show the motion estimation capability for periodic (Fig. 6) and linear drifting (Fig. 7) respiration in two patients with neuroendocrine tumors. Motion estimates obtained from shorter 1 min portions of the scan (i.e. corresponding to ~25x undersampling), are consistent for cyclic motion. Deep learning based registrations are superior to conventional image registration (NiftyReg). Flows of the proposed LAPNet are in accordance with reference flows. For linear drifting respiration, FlowNet-S and NiftyReg fail to capture deformations accurately. LAPNet produced in this scenario reliable results.

The depicted cases in Fig. 4-7 show the observed performance variations amongst test subjects. No outlier or failed cases were observed in the test cohort with similar performance amongst subjects.

Non-rigid registration with LAPNet required ~12 hours in training whereas inference required only ~18s. FlowNet-S required ~2.5 hours of training and ~12s in inference.

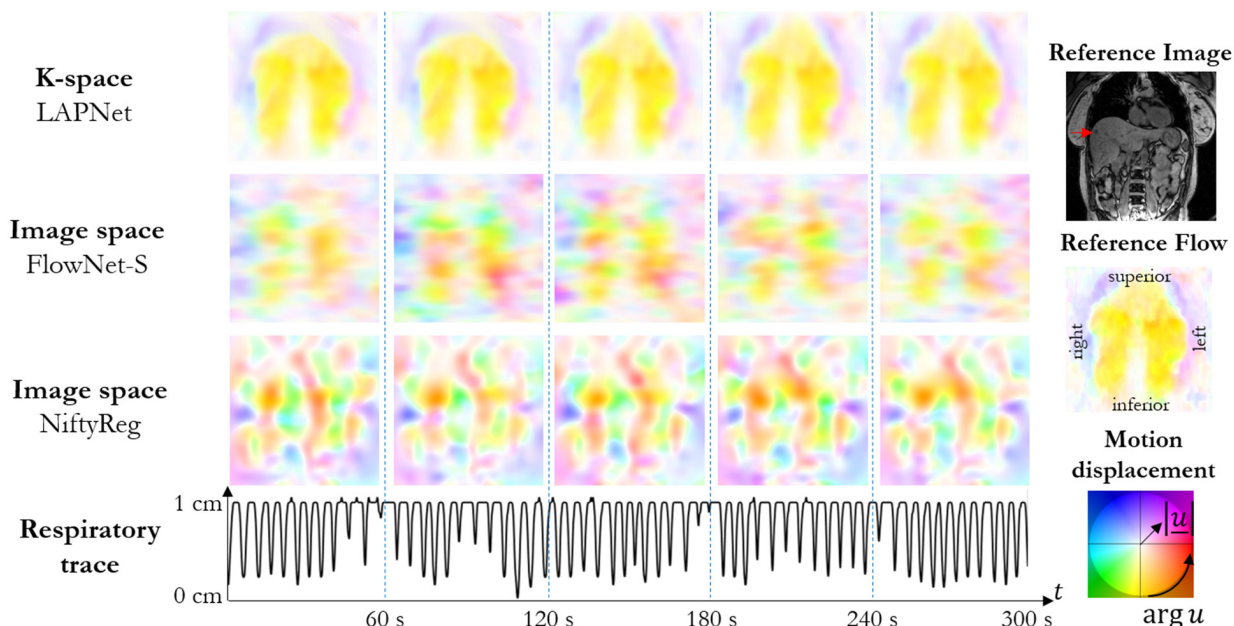

Fig. 6: Respiratory non-rigid motion estimation in a patient with neuroendocrine tumor and liver metastastis (pointed out by red arrow). Motion displacement is estimated by the proposed LAPNet in k-space in comparison to image-based non-rigid registration by FlowNet-S (deep learning) and NiftyReg (cubic B-Splines). Motion is estimated from 1 min parts of the scan, corresponding to a ~25x undersampling. Reference flows depict the imageLAP registration from the complete 5 min scan. Cyclic respiration with 1 cm super-inferior displacement was observed.

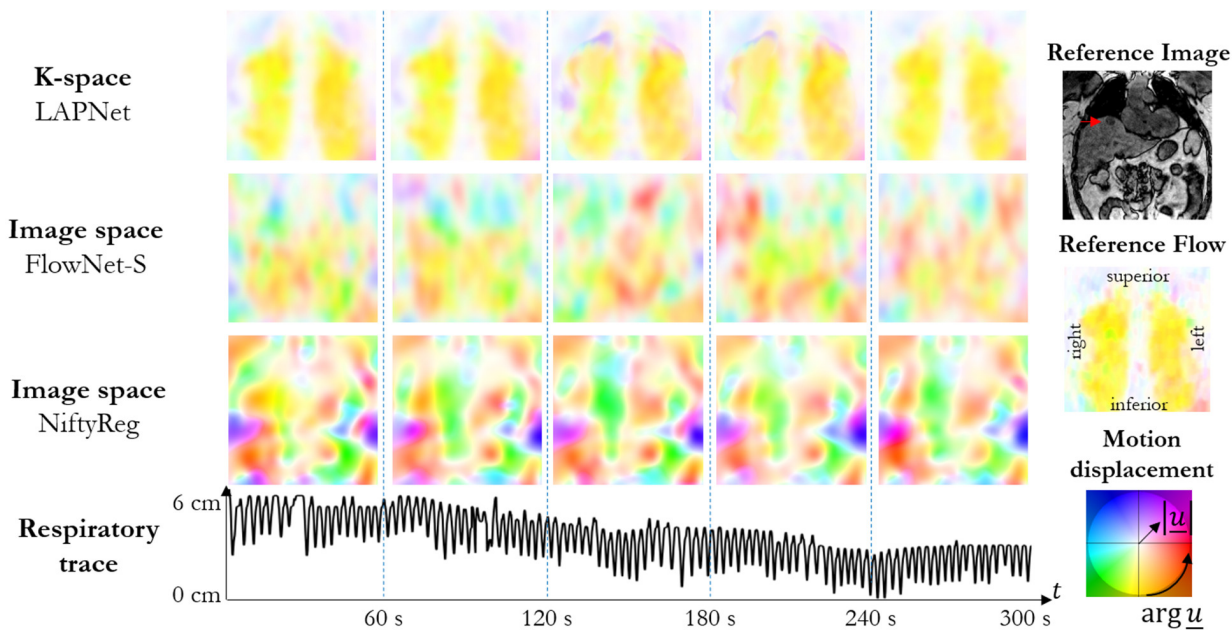

Fig. 7: Respiratory non-rigid motion estimation in a patient with neuroendocrine tumor and liver metastastis (pointed out by red arrow). Motion displacement is estimated by the proposed LAPNet in k-space in comparison to image-based non-rigid registration by FlowNet-S (deep learning) and NiftyReg (cubic B-Splines). Motion is estimated from 1 min parts of the scan, corresponding to a ~25x undersampling. Reference flows depict the imageLAP registration from the complete 5 min scan. Linear drifting respiration with ~3 cm super-inferior displacement was observed.

Registrations with conventional methods were performed on average in 23 ± 4s for imageLAP and 83 ± 28s for NiftyReg.

## V. DISCUSSION

In this work, we propose a deep learning network based non-rigid registration in k-space. The proposed LAPNet was compared against image-based non-rigid registration methods in healthy subjects and patients for *fully-sampled* acquisitions and two undersampling strategies (*vdPD* and *center*). A throughout consistent and superior performance of the proposed LAPNet was found. Non-rigid motion estimation is feasible in k-space which generalizes well for highly accelerated cases.

Data-driven learning of image registration (LAPNet, FlowNet-S) outperformed conventional image registrations (imageLAP, NiftyReg) for accelerated acquisitions, although being trained in a supervised manner on imageLAP derived flows. The LAPNet, i.e. k-space registration, showed a more consistent and improved performance over image-based registration (FlowNet-S) for different acceleration factors.

A comparison of the proposed LAPNet to image registration methods that are more robust towards undersampled aliasing and blurring would be desirable. However, registration methods in this field are primarily paired up or interleaved into an image reconstruction [28-35]. Examination of sole impact and effect of image registration becomes hereby difficult. We therefore used the image-based conventional and deep learning image registrations as comparison.

The influence of the receptive field for LAPNet and FlowNet-S were investigated. In LAPNet, an empirically optimized size $W$ was determined which provided a good trade-off between computational time and accuracy. For increasing receptive field size, non-rigid motion estimation was less accurate and overall performance dropped. If receptive field was chosen too small, the local neighbourhood was not contributing to the non-rigid motion estimation resulting in non-smooth flows. A good trade-off was found for $W = 33$ and filter kernel sizes. In contrast to imageLAP, maximum possible displacement that can be registered with LAPNet was not confined by cubic size $W$. Here, the question for maximum possible displacement that can be registered corresponds to maximum possible perceived phase changes. Furthermore, it is expected that data-driven deep learning registrations can benefit from learning various motion patterns and displacement amplitudes. In the examined cases, motion of varying amplitude was seen for which LAPNet provided good registrations. Future investigations are required to examine registration accuracy limits.

We observed that for the same acceleration factor the LAPNet seems to handle motion estimation in the presence of aliasing artifacts (image space) better as obtained with *vdPD* sampling than from low-resolution data as in *center* sampling (see Fig. 1, 2). For high acceleration factors (>15x) it seems to be beneficial to include some high-frequency samples in the registration. In these accelerations, more low-frequency samples are contributing to the estimation, suggesting that there might be an optimal amount and distribution of low-frequency and high-frequency samples which contribute to the overall registration. It is hence conceivable training a network to identify the most significant k-space samples contributing to the registration which is related to works of adaptive sampling for image reconstruction [71].

Qualitatively, smoother flow estimations were observed from *center* sampling in the test cohort. Especially FlowNet-S was operating better under blurring than aliasing artifacts. While quantitative metrics are often used for driving the conventional image registration tasks, the voxel-wise measures do not necessarily reflect the correction of motion [72], but provide a good indicator of how well each of the respective methods perform in relation to each other.

Spatial alignment between flows and undersampled data was achieved by retrospective undersampling in training. In order to account in the supervised training for changing and subject-specific respiratory motion patterns, augmentation of real flows was performed. Bilinear interpolation can introduce interpolation errors but was found to be sufficient and least computationally demanding for training. Testing was conducted on prospectively undersampled data with varying motion patterns. Consistent and superior performance of the proposed LAPNet was found for cases ranging from mild (up to 1.5 cm liver dome displacement) to strong motion (up to 6 cm) with linear drifting respiration. However, future investigation on larger cohorts is warranted with strongly varying motion patterns.

We acknowledge further limitations of this work. The proposed LAPNet was trained and tested for a single imaging sequence with two different sampling strategies in a coil-combined setting for 2D k-space input. In the future, we want to extend LAPNet for multi-coil 3D k-space input and investigate its generalizability for different imaging sequences and sampling trajectories, e.g. radial sampling, as well as applications. An empirically optimized input size was set, but a multi-resolution approach is conceivable and needs to be investigated in the future. The trained filters are not restricted to have all-pass characteristic but approximate their behaviour which needs to be further investigated in the future. A supervised learning was used which may be biased by the image-based LAP reference flow, potentially also limiting FlowNet-S capabilities. Future work will investigate on self-supervised and unsupervised learning together with other suitable dissimilarity measures in Fourier domain.

## VI. CONCLUSION

A deep-learning based non-rigid registration method, LAPNet, which can be directly performed in the acquired k-space domain is proposed. Results indicated improved performance of LAPNet in comparison to image-based registration approaches for high acceleration factors. LAPNet showed consistent performance throughout different sampling trajectories and acceleration factors. It thus enables non-rigid motion registration from highly accelerated acquisitions.

## ACKNOWLEDGMENT

The authors would like to thank Brigitte Gückel for study coordination as well as Carsten Groeper and Gerd Zeger in assisting the data acquisition. We thank Kerstin Hammernik for helpful discussions.